\documentstyle[12pt,epsfig]{article}
\def\be{\begin{equation}}
\def\ee{\end{equation}}
\def\ba{\begin{array}}
\def\ea{\end{array}}

\begin{document}
\title{\bf Centrifugal-Barrier Effects and Determination of
    the Interaction Radius }
\author{{Ning Wu }
\thanks{email address: wuning@ihep.ac.cn}
\\
\\
{\small Institute of High Energy Physics, P.O.Box 918-1,
Beijing 100039, P.R.China}}
\maketitle

\begin{abstract}

The interaction radius of a resonance is an important physical quantity
to describe the structure of
a resonance. But, for a long time, physicists do not
find a reliable way to  measure the magnitude of
the interaction radius of a resonance. In this paper, a method is proposed
to measure the interaction radius in physics analysis.
 It is found that the centrifugal
barrier effects have great influence to physical results obtained in the
PWA fit, and the interaction radius of some resonances can be well
measured in the fit.

\end{abstract}

\Roman{section}

\section{Introduction}
\setcounter{equation}{0}

Partial Wave Analysis(PWA) now is widely used in physics analysis
of high energy experimental physics. As an effective method,
by analyzing information of final state particles, it is used to
search new resonances in a complicated spectrum, and to
determine mass, width, branching ratio and spin-parity of an intermediate
resonance. In the traditional PWA analysis, the centrifugal barrier effects
are considered.  It is known that, after a resonance
decays into final state particles, the outward-going wave is being reflected
inward by the centrifugal barrier. Therefore, the radial probability density at
infinity is the centrifugal  barrier factor $|BF(L,kR)|^2$ times the
radial probability density at the interaction radius $R$.
\\

The centrifugal barrier effects are studied in literatures \cite{01,02}. Basic results
of the papers are that the shape and the partial decay width of a
resonance are affected by the centrifugal barrier effects. The centrifugal barrier
factor $BF(L,kR)$ has two parameters, the orbital angular momentum
quantum number $L$ and the interaction radius $R$. It shows that the interaction
radii range from 0.25 to 0.75 F for the meson resonance decays and from
0.1 to 1 F for the baryon decays.
\\

It is conventionally believed that the influence of the centrifugal barrier effects
to physics analysis is small. Therefore, the centrifugal barrier effects are not
considered in some physics analysis, which corresponds to the case that
$BF(L,kR)=1$ or $R=\infty$. In partial wave analysis, the centrifugal barrier
effects are considered, but the interaction radius $R$ is set to 1 F for all
resonances\cite{03,04,05,06,07,08}. In this paper, the influence of the centrifugal
barrier effects to physics analysis is studied, and a method is proposed
to measure the interaction radius in physics analysis. First, the decay
amplitude used in the physics analysis is given. Then, the problem that
how the resonance's line shape is affected by the centrifugal barrier effects
is studied.  Finally, we will discuss how to measure the
interaction radius in a PWA fit.
\\

\section{Centrifugal Barrier Factor}
\setcounter{equation}{0}

Let's consider a two body decay $a \to b + c$.In the center
of mass system of the mother particle $a$, the magnitude of the
momentum of the particle $b$ is $k$, and the orbital angular momentum
quantum number of the daughter particle system is $l$. Suppose that
the interaction potential $V(r)$ is almost zero when
$r \geq R_0$, where $R_0$ is the interaction radius of the
mother particle $a$.
The wave equation which holds outside the interaction
radius $R_0$ is
\be \label{01}
R'' + \frac{2}{r} R' + (k^2 - \frac{l(l+1)}{r^2})R =0,
\ee
where $R$ is the radial wave function, and
$k=\sqrt{\frac{2 \mu E}{\hbar^2}}$ is the semiclassical
impact parameter. It has two independent solutions,
$h_l^{(1)}(\rho)$ and $h_l^{(2)}(\rho)$, where
$\rho = k r$. The probability density
at $r$ is proportional to $|\rho R(\rho)|^2$. Define
\be \label{02}
BF(l,\rho) = \frac{1}{|\rho h_l^{(1)}(\rho)|}.
\ee
Then, the probability at infinity is the probability at $R_0$
times $|BF(l,\rho_0)|^2$ with $\rho_0 = k R_0$.
Some frequently used barrier factors are
\be \label{03}
BF(0,\rho_0) = 1,
\ee

\be \label{04}
BF(1,\rho_0) =
\sqrt{\frac{2 z}{z+1}},
\ee

\be \label{05}
BF(2,\rho_0) =
\sqrt{\frac{13 z^2}{z^2 + 3 z + 9}},
\ee

\be \label {06}
BF(3,\rho_0) =
\sqrt{\frac{277 z^3}{z^3 + 6 z^2 + 45 z + 225}},
\ee

\be \label{07}
BF(4,\rho_0) =
\sqrt{\frac{12746 z^4}{z^4 + 10 z^3 + 135 z^2 + 1575 z + 11025}},
\ee

\be \label{08}
BF(5,\rho_0) =
\sqrt{\frac{998881 z^5}{z^5 + 15 z^4 + 315 z^3 + 6300 z^2 + 99225 z + 893025}},
\ee

\be \label{09}
BF(6,\rho_0) =
\sqrt{\frac{118394977 z^6}{z^6 + 21 z^5 + 630 z^4 + 18900 z^3 + 496125 z^2 + 9823275 z + 108056025}},
\ee
where $z=\rho_0^2$. Figure 1 shows the shapes of $|BF(L,kR)|^2$ with
L=2,4,6 and 8. If $L=0$, $BF(0,kR)=1$, which means that the line shape of
a scalar resonance is not affected by the centrifugal barrier effects.
 In this figure, all curves are normalized
to 1 when the momentum $k$ approaches infinity. It could be seen
that the shapes of these curves change greatly when the momentum $k$ is in
the  region 0-2 GeV, which is just in the momentum region of
final state particles produced in the decays of
most mesons or hadrons. Different shapes of barrier factors  give out
different line shapes of resonances. If the centrifugal barrier effects are
not correctly considered, the line shape of a resonance given by it
will be quite different from the one with  the centrifugal barrier factors
correctly considered, which will cause large errors in the measurement
of mass, width and branching ratio of a resonance. If the momentum of the
daughter particle in the center of mass system of the mother
particle is greater than 3 GeV, the influence of the
centrifugal barrier effects to such kind of decay will be
small. In this case, the mass of mother particle must be greater
than 6 GeV. Therefore, if we study the decay of a low mass resonance,
the influence of the centrifugal barrier effects must be
considered, otherwise, the systematic uncertainty caused by it
will be expected to be quite large.
\\

\begin{figure}[htbp]
\vspace{-0.5in}
\begin{center}
{\mbox{\epsfig{file=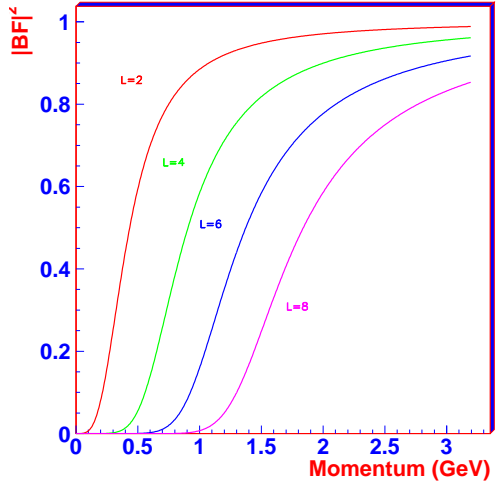,height=2.8in,width=2.8in}}}
\end{center}
\caption[]{The shape of $|BF(L,kR)|^2$ with
L=2,4,6 and 8. The interaction radius $R_0$ is set to 1 fm in this figure.
    }
\label{k01}
\end{figure}

\section{Decay Amplitude}

Now, let's study the influence of the centrifugal barrier effects
to the resonance's line shape. We will study how the resonance's line
shape is changed by the centrifugal barrier effects. The barrier
factor of a resonance decay not only affects its own line shape, but
also affects the line shapes of its daughter particles. Therefore,
for a sequential decay, the line shape of an intermediate
resonance is affected by more than one barrier factor.
As an example, let's consider the sequential decay
$J/\psi \to \omega X \to \omega \pi \pi$. The spin-parity
of the resonance $X$ is $J^{PC}$. In this sequential decay, there
are two vertexes, one is the decay $J/\psi \to \omega X$,
and another is the decay $X \to \pi \pi$. The decay amplitude
of the sequential decay is\cite{09,03,04,05,06,07,08}
\be \label{10}
M \sim F^{(1) 1}_{\lambda \nu} D^{1* }_{M (\lambda - \nu)}(\phi_1,\theta_1,0)
\cdot BW \cdot F^{(2)J}_{00} D^{J*}_{\lambda 0}(\phi_2, \theta_2, 0),
\ee
where $F^{(1)1}_{\lambda \nu}$ is the helicity amplitude of the decay
$J/\psi \to \omega X$, $BW$ is the Breit-Wigner function
of the resonance $X$, and $F^{(2)J}_{00}$ is the helicity amplitude
of the decay $X \to \pi \pi$.
\\

If the spin-parity of the resonance $X$ is $0^+$, then the independent
helicity amplitudes are
\be \label{11}
F^{(1) 1}_{01} =
g_1 - \frac{g_2}{3}  BF(2,k_1 R_1),
\ee

\be \label{12}
F^{(1) 1}_{00} =
\gamma_{\sigma} \left (g_1 + \frac{2 g_2}{3}  BF(2,k_1 R_1) \right ),
\ee

\be \label{13}
F^{(2) 0}_{00} =
g_3,
\ee
where $g_1$, $g_2$ and $g_3$ are scalars, $\gamma_{\sigma}$ is the rapidity
of  $\omega$ particle, $k_1$ is the relative momentum of $\omega$
particle in the center of mass system of $J/\psi$,
 and $R_1$ is the interaction radius of $J/\psi$.
If the spin-parity of the resonance $X$ is $2^+$, then the independent
helicity amplitudes are
\be \label{14}
\ba{rcl}
F^{(1) 1}_{21} & = &
\frac{g_1}{10} (-7 -3 \gamma_s \gamma_{\sigma}
    + 3 \beta_s \beta_{\sigma} \gamma_s \gamma_{\sigma})
+ \frac{g_2}{30}  (7 - 6 \gamma_s \gamma_{\sigma})\cdot BF(2, k_1 R_1)\\
&&\\
&& + \frac{g_3}{18}  (1+2 \gamma_s \gamma_{\sigma}) \cdot BF(2, k_1 R_1)
+ \frac{2 g_4}{45}  (1 + 2 \gamma_s \gamma_{\sigma}) \cdot BF(2, k_1 R_1)\\
&&\\
&& - \frac{2 g_5}{525}  (3 + 4 \gamma^2_s + 8 \gamma_s \gamma_{\sigma}) \cdot BF(4, k_1 R_1),
\ea
\ee

\be \label{15}
\ba{rcl}
F^{(1) 1}_{11} & = &
\frac{g_1}{2 \sqrt{2}}  \gamma_s (-3 -2 \gamma_s \gamma_{\sigma}
    + 2 \beta_s \beta_{\sigma} \gamma_s \gamma_{\sigma})
+ \frac{g_2}{15 \sqrt{2}}  \gamma_s (3 - 4 \gamma_s \gamma_{\sigma}) \cdot BF(2, k_1 R_1)\\
&&\\
&& + \frac{\sqrt{2} g_4}{15} \gamma_s (1 + 2 \gamma_s \gamma_{\sigma}) \cdot BF(2, k_1 R_1)\\
&&\\
&&  + \frac{4 \sqrt{2} g_5}{175}  (2 \gamma_s + \gamma_{\sigma}
    + 2 \gamma^2_s \gamma_{\sigma}) \cdot BF(4, k_1 R_1),
\ea
\ee

\be \label{16}
\ba{rcl}
F^{(1) 1}_{10} & = &
\frac{g_1}{10 \sqrt{2}} (7 + 3 \gamma_s \gamma_{\sigma}
    - 3 \beta_s \beta_{\sigma} \gamma_s \gamma_{\sigma})
- \frac{g_2}{30 \sqrt{2}}  (7 - 6 \gamma_s \gamma_{\sigma})\cdot BF(2, k_1 R_1)\\
&&\\
&& + \frac{g_3}{18 \sqrt{2}}  (1+2 \gamma_s \gamma_{\sigma}) \cdot BF(2, k_1 R_1)
+ \frac{8 g_4}{45 \sqrt{2}}  (1 + 2 \gamma_s \gamma_{\sigma}) \cdot BF(2, k_1 R_1)\\
&&\\
&& - \frac{8 g_5}{525 \sqrt{2}}  (3 + 4 \gamma^2_s + 8 \gamma_s \gamma_{\sigma}) \cdot BF(4, k_1 R_1),
\ea
\ee

\be \label{17}
\ba{rcl}
F^{(1) 1}_{01} & = &
\frac{g_1}{10 \sqrt{6}} (-7 - 3 \gamma_s \gamma_{\sigma}
    + 3 \beta_s \beta_{\sigma} \gamma_s \gamma_{\sigma})
+ \frac{g_2}{30 \sqrt{6}}  (7 - 6 \gamma_s \gamma_{\sigma})\cdot BF(2, k_1 R_1)\\
&&\\
&& - \frac{g_3}{6 \sqrt{6}}  (1+2 \gamma_s \gamma_{\sigma}) \cdot BF(2, k_1 R_1)
+ \frac{4 g_4}{15 \sqrt{6}}  (1 + 2 \gamma_s \gamma_{\sigma}) \cdot BF(2, k_1 R_1)\\
&&\\
&& - \frac{4 g_5}{175 \sqrt{6}}  (3 + 4 \gamma^2_s + 8 \gamma_s \gamma_{\sigma}) \cdot BF(4, k_1 R_1),
\ea
\ee

\be \label{18}
\ba{rcl}
F^{(1) 1}_{00} & = &
\frac{2 g_1}{5 \sqrt{6}}  \gamma_s (3 + 2 \gamma_s \gamma_{\sigma}
    - 2 \beta_s \beta_{\sigma} \gamma_s \gamma_{\sigma})
- \frac{2 g_2}{15 \sqrt{6}}  \gamma_s (3 - 4 \gamma_s \gamma_{\sigma}) \cdot BF(2, k_1 R_1)\\
&&\\
&& + \frac{2 g_4}{5 \sqrt{6}} \gamma_s (1 + 2 \gamma_s \gamma_{\sigma}) \cdot BF(2, k_1 R_1)\\
&&\\
&&  + \frac{4 \sqrt{6} g_5}{175}  (2 \gamma_s + \gamma_{\sigma}
    + 2 \gamma^2_s \gamma_{\sigma}) \cdot BF(4, k_1 R_1).
\ea
\ee

\be \label{19}
F^{(2) 2}_{00} =
g_6 \cdot BF(2, k_2 R_2),
\ee
where $g_1$, $g_2$, $g_3$, $g_4$, $g_5$ and $g_6$ are scalars,
$\gamma_s$ and $\gamma_{\sigma}$ are the rapidities
of  $X$ and $\omega$ respectively,
$\beta_s$ and $\beta_{\sigma}$ are the velocities
of  $X$ and $\omega$  respectively,
$k_1$ is the relative momentum of $\omega$
particle in the center of mass system of $J/\psi$,
 $R_1$ is the interaction radius of $J/\psi$,
 $k_2$ is the relative momentum of $\pi$
particle in the center of mass system of $X$,
and  $R_2$ is the interaction radius of resonance $X$.
If the spin-parity of the resonance $X$ is $4^+$, then the independent
helicity amplitudes are
\be \label{20}
\ba{rcl}
F^{(1) 1}_{21} & = &
\frac{g_1}{36 \sqrt{7}} (1 + 2 \gamma_s^2)
    (-7 -5 \gamma_s \gamma_{\sigma}
    + 5 \beta_s \beta_{\sigma} \gamma_s \gamma_{\sigma}) \cdot BF(2, k_1 R_1)\\
    &&\\
&& + \frac{g_2}{420 \sqrt{7}}  (21 + 84 \gamma_s^2 - 60 \gamma_s \gamma_{\sigma}
    - 40 \gamma_s^3 \gamma_{\sigma}) \cdot BF(4, k_1 R_1)\\
&&\\
&& + \frac{3 g_3}{700 \sqrt{7}}  (3 + 12 \gamma_s^2
    + 12 \gamma_s \gamma_{\sigma}  + 8 \gamma_s^3 \gamma_{\sigma}) \cdot BF(4, k_1 R_1)\\
&&\\
&& + \frac{4 g_4}{525 \sqrt{7}}  (3 + 12 \gamma_s^2
    + 12 \gamma_s \gamma_{\sigma}  + 8 \gamma_s^3 \gamma_{\sigma}) \cdot BF(4, k_1 R_1)\\
&&\\
&& - \frac{4 g_5}{3465 \sqrt{7}}  (5 + 36 \gamma_s^2 + 8 \gamma_s^4
    + 24 \gamma_s \gamma_{\sigma}  + 32 \gamma_s^3 \gamma_{\sigma}) \cdot BF(6, k_1 R_1),
\ea
\ee

\be \label{21}
\ba{rcl}
F^{(1) 1}_{11} & = &
\frac{g_1}{27 \sqrt{7}} \gamma_s (1 + 2 \gamma_s^2)
    (-5 -4 \gamma_s \gamma_{\sigma}
    + 4 \beta_s \beta_{\sigma} \gamma_s \gamma_{\sigma}) \cdot BF(2, k_1 R_1)\\
    &&\\
&& + \frac{g_2}{315 \sqrt{7}} \gamma_s (15 + 60 \gamma_s^2 - 48 \gamma_s \gamma_{\sigma}
    - 32 \gamma_s^3 \gamma_{\sigma}) \cdot BF(4, k_1 R_1)\\
&&\\
&& + \frac{4 g_4}{315 \sqrt{7}} \gamma_s (3 + 12 \gamma_s^2
    + 12 \gamma_s \gamma_{\sigma}  + 8 \gamma_s^3 \gamma_{\sigma}) \cdot BF(4, k_1 R_1)\\
&&\\
&& + \frac{8 g_5}{2079 \sqrt{7}}  (12 \gamma_s + 16 \gamma_s^3 + 3 \gamma_{\sigma}
    + 24 \gamma_s^2 \gamma_{\sigma}  + 8 \gamma_s^4 \gamma_{\sigma}) \cdot BF(6, k_1 R_1),
\ea
\ee

\be \label{22}
\ba{rcl}
F^{(1) 1}_{10} & = &
\frac{g_1}{36 \sqrt{7}} (1 + 2 \gamma_s^2)
    (7 + 5 \gamma_s \gamma_{\sigma}
    - 5 \beta_s \beta_{\sigma} \gamma_s \gamma_{\sigma}) \cdot BF(2, k_1 R_1)\\
    &&\\
&& - \frac{g_2}{420 \sqrt{7}}  (21 + 84 \gamma_s^2 - 60 \gamma_s \gamma_{\sigma}
    - 40 \gamma_s^3 \gamma_{\sigma}) \cdot BF(4, k_1 R_1)\\
&&\\
&& + \frac{ g_3}{700 \sqrt{7}}  (3 + 12 \gamma_s^2
    + 12 \gamma_s \gamma_{\sigma}  + 8 \gamma_s^3 \gamma_{\sigma}) \cdot BF(4, k_1 R_1)\\
&&\\
&& + \frac{8 g_4}{525 \sqrt{7}}  (3 + 12 \gamma_s^2
    + 12 \gamma_s \gamma_{\sigma}  + 8 \gamma_s^3 \gamma_{\sigma}) \cdot BF(4, k_1 R_1)\\
&&\\
&& - \frac{8 g_5}{3465 \sqrt{7}}  (5 + 36 \gamma_s^2 + 8 \gamma_s^4
    + 24 \gamma_s \gamma_{\sigma}  + 32 \gamma_s^3 \gamma_{\sigma}) \cdot BF(6, k_1 R_1),
\ea
\ee

\be \label{23}
\ba{rcl}
F^{(1) 1}_{01} & = &
\frac{g_1}{18 \sqrt{70}} (1 + 2 \gamma_s^2)
    (-7 -5 \gamma_s \gamma_{\sigma}
    + 5 \beta_s \beta_{\sigma} \gamma_s \gamma_{\sigma}) \cdot BF(2, k_1 R_1)\\
    &&\\
&& + \frac{g_2}{210 \sqrt{70}}  (21 + 84 \gamma_s^2 - 60 \gamma_s \gamma_{\sigma}
    - 40 \gamma_s^3 \gamma_{\sigma}) \cdot BF(4, k_1 R_1)\\
&&\\
&& - \frac{ g_3}{70 \sqrt{70}}  (3 + 12 \gamma_s^2
    + 12 \gamma_s \gamma_{\sigma}  + 8 \gamma_s^3 \gamma_{\sigma}) \cdot BF(4, k_1 R_1)\\
&&\\
&& + \frac{4 g_4}{105 \sqrt{70}}  (3 + 12 \gamma_s^2
    + 12 \gamma_s \gamma_{\sigma}  + 8 \gamma_s^3 \gamma_{\sigma}) \cdot BF(4, k_1 R_1)\\
&&\\
&& - \frac{4 g_5}{693 \sqrt{70}}  (5 + 36 \gamma_s^2 + 8 \gamma_s^4
    + 24 \gamma_s \gamma_{\sigma}  + 32 \gamma_s^3 \gamma_{\sigma}) \cdot BF(6, k_1 R_1),
\ea
\ee

\be \label{24}
\ba{rcl}
F^{(1) 1}_{00} & = &
\frac{4 g_1}{27 \sqrt{70}} \gamma_s (1 + 2 \gamma_s^2)
    (5 +4 \gamma_s \gamma_{\sigma}
    - 4 \beta_s \beta_{\sigma} \gamma_s \gamma_{\sigma}) \cdot BF(2, k_1 R_1)\\
    &&\\
&& - \frac{4 g_2}{315 \sqrt{70}} \gamma_s (15 + 60 \gamma_s^2 - 48 \gamma_s \gamma_{\sigma}
    - 32 \gamma_s^3 \gamma_{\sigma}) \cdot BF(4, k_1 R_1)\\
&&\\
&& + \frac{4 g_4}{63 \sqrt{70}} \gamma_s (3 + 12 \gamma_s^2
    + 12 \gamma_s \gamma_{\sigma}  + 8 \gamma_s^3 \gamma_{\sigma}) \cdot BF(4, k_1 R_1)\\
&&\\
&& + \frac{40 g_5}{2079 \sqrt{70}}  (12 \gamma_s + 16 \gamma_s^3 + 3 \gamma_{\sigma}
    + 24 \gamma_s^2 \gamma_{\sigma}  + 8 \gamma_s^4 \gamma_{\sigma}) \cdot BF(6, k_1 R_1),
\ea
\ee

\be \label{25}
F^{(2) 4}_{00} =
g_6 \cdot BF(4, k_2 R_2).
\ee
\\

In the traditional PWA analysis, the interaction radii $R_1$ and $R_2$ are
set to 1 F. It is obvious that such treatment is not reasonable, for different
resonance has different interaction radius. If the systematic uncertainty caused
by it is small enough, the treatment is acceptable. Otherwise, we must
find a way to determine  the interaction radius in physics analysis.
Next, we will study how the resonance's line
shape is affected by the interaction radius, and how to determine the interaction radius
in physics analysis.
\\

\section{Resonance's Line Shape}

In order to know how the resonance's line shape is affected by
the interaction radius, we consider
a simplified model, that is, only the centrifugal barrier
effects are considered in the helicity amplitudes
$F^{(1)1}_{\lambda \nu}$ and $F^{(2)J}_{00}$. In this simplified
model, the decay amplitude of the sequential decay is
\be \label{26}
M \sim BF^{(1)}(L,k_1 R_1)
\cdot BW \cdot BF^{(2)}(J,k_2 R_2),
\ee
where $k_1$ and $k_2$ are relative momenta of daughter
particles in the center of mass system of the first and the second
decay vertexes respectively, and $R_1$ and $R_2$ are the
corresponding interaction radiuses of two mother particles.
The absolute square of the decay amplitude of the sequential decay
gives out the line shape of the  resonance $X$.
It could be seen that  the line shape of the resonance $X$ is affected by
two independent centrifugal barrier effects, one is the
centrifugal barrier effect of the decay of the resonance
$X$ itself, another is the centrifugal barrier effect of
the decay of the mother particle whose decay produces
resonance $X$. Next, we will study these  two barrier effects
separately.
\\

$BF^{(2)}(J,k_2 R_2)$ is the barrier factor of the decay of
the resonance $X$. In order to know how the line shape is affected by
$BF^{(2)}(J,k_2 R_2)$, we study the shape of
$|BW(s,m,\Gamma) \cdot BF^{(2)}(J,k_2 R_2)|^2$.
Figure 2 shows the resonance's line shape after the
influence of $BF^{(2)}(J,k_2 R_2)$ ( J=0,2,4, and 6) is considered.
The curves in figure 2 show the shapes of
$|BW(s,m,\Gamma) \cdot BF^{(2)}(J,k_2 R_2)|^2$.
In this model, the mass and the width of the resonance $X$ are
set to 1.27 GeV and 0.185 GeV respectively, and
$Pr = \frac{1}{R}$ is set to 0.1973 GeV, which corresponds
to 1 F interaction radius. In this figure, we could see that the
influence of the centrifugal barrier effects to the
resonance's line shape is that the lower part
of the resonance's line shape is suppressed and the higher part
is enhanced. The total effect of the influence from the centrifugal
barrier effects is to make the resonance's line shape
unsymmetric. And this influence becomes larger when the orbital angular
momentum quantum number $J$ is bigger. In this figure,
the influence from the centrifugal barrier effects to the
resonance's line shape is quite large. Therefore, it is expected that
the systematic errors caused by it will be large if it is not correctly
considered in the analysis. In some physics analysis, the centrifugal
barrier effects are completely not considered, physics
results obtained by it will have large systematic errors. According to
a test on BESII data, the systematic errors to the mass and the width of
a resonance caused by this reason are generally about 20-100 MeV and
the relative uncertainties of branching ratio is about 30-50 \%.
\\

\begin{figure}[htbp]
\vspace{-0.5in}
\begin{center}
{\mbox{\epsfig{file=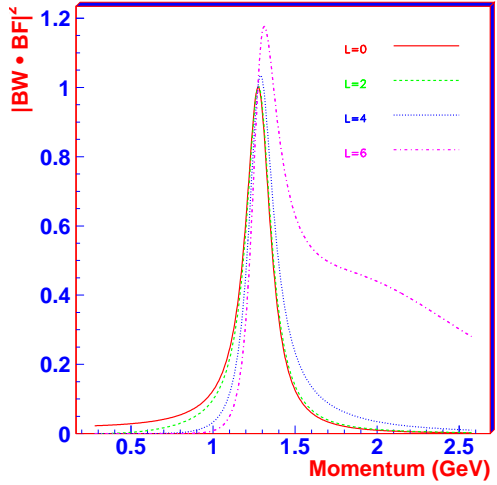,height=2.8in,width=2.8in}}}
\end{center}
\caption[]{The shape of $|BW(s,m,\Gamma) \cdot BF^{(2)}(J,k_2 R_2)|^2$ with
J=0,2,4, and 6. The interaction radius $R_0$ is set to 1 F in this figure.
    }
\label{k02}
\end{figure}

Another barrier factor that affects the line shape of the resonance $X$ is that
of the $J/\psi$ decay, that is, $BF^{(1)}(L,k_1 R_1)$. In order to
know its influence to the resonance's line shape, let's study the shape
of $|BF^{(1)}(L,k_1 R_1) \cdot BW(s,M,\Gamma)|^2 $. Figure 3 shows
the curves of $|BF^{(1)}(L,k_1 R_1) \cdot BW(s,M,\Gamma)|^2 $ with
L=0,2,4, and 6. Contrary to the barrier factor $BF^{(2)}(J,k_2 R_2)$,
$BF^{(1)}(L,k_1 R_1)$ suppresses the higher part of the resonance's
line shape and enhances its lower part. Quantitatively speaking, the influence
is relatively much smaller than that of $BF^{(2)}(J,k_2 R_2)$.
\\

\begin{figure}[htbp]
\vspace{-0.5in}
\begin{center}
{\mbox{\epsfig{file=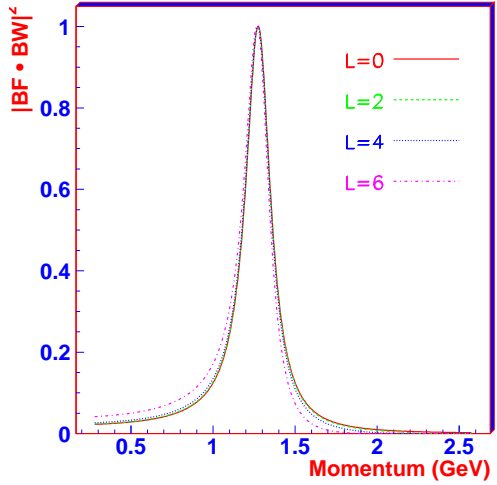,height=2.8in,width=2.8in}}}
\end{center}
\caption[]{The shape of $|BF^{(1)}(L,k_1 R_1) \cdot BW(s,M,\Gamma)|^2$ with
L=0,2,4, and 6. The interaction radius $R_0$ is set to 1 F in this figure.
    }
\label{k03}
\end{figure}

In most partial wave analysis, the centrifugal barrier effects
are considered and the interaction radius $R$ is set to 1 F for
all resonances.
According to the literature \cite{02}, the interaction radius $R$ is not
exact 1 F. The interaction radii range from 0.25 to 0.75 F
for the meson resonance decays and from 0.5 to 1 F for
the baryon decays. Uncertainties on the interaction radius $R$
will cause systematic uncertainties in physics analysis.
If different values of the interaction radius do not cause obvious
changes on the resonance's line shape, the systematic uncertainties
caused by it will be small. In this case, no matter what values
the interaction radius is, final physical results are acceptable.
But if resonance's line shape is sensitive to the value of the interaction
radius, we can not fix the value of the interaction radius arbitrarily
in physics analysis. Now, we  study the influence of
the interaction radius to the resonance's line shape.
We consider the decay of a spin-2 particle,
and neglect the centrifugal barrier effects of
the $J/\psi$ decay. In this model, the decay amplitude  is
\be \label{27}
M \sim
 BW \cdot BF^{(2)}(2,k_2 R_2).
\ee
The line shape of the spin-2 particle is shown in Figure 4. Dashed line
is the line shape of the spin-2 particle when its interaction radius
is set to 1 F. Solid line, dotted line and dot-dashed line are the corresponding
line shapes when its interaction radius is set to 10/3 F, 0.5 F and
0.2 F respectively. From this figure, we could clearly see that the
influence of the interaction radius to the resonance's line shape
is obvious and quite large. If the resonance is a scalar particle, four
curves will coincide, for $BF^{(2)}(0,k_2 R_2)=1$. In other words,
the line shape of a scalar resonance is not affected by its centrifugal
barrier factor. If the spin of the resonance is 4, the influence of the
centrifugal barrier factor will become larger. Figure 5 shows the line shapes
of a spin-4 resonance with interaction radius  0.2 F, 0.5 F, 1 F and 10/3 F
respectively. From these two figures, our intuitive impression is that
the influence of the centrifugal barrier effects is to make the resonance's
line shape unsymmetric, which provides us a method to measure the exact
value of the resonance's interaction radius.
\\

\begin{figure}[htbp]
\vspace{-0.5in}
\begin{center}
{\mbox{\epsfig{file=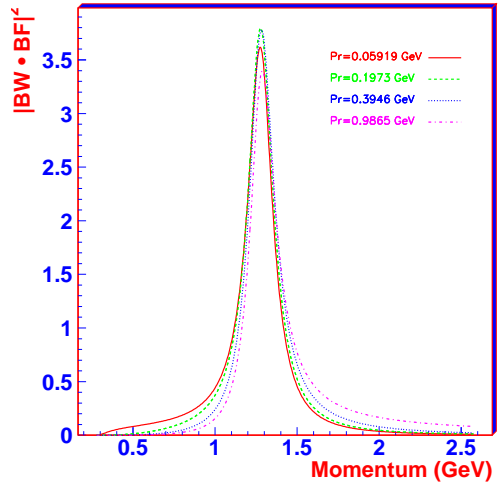,height=2.8in,width=2.8in}}}
\end{center}
\caption[]{The shape of $|BW(s,M,\Gamma) \cdot BF^{(2)}(2,k_2 R_2)|^2$ with
interaction radius $R_0$ 0.2 F, 0.5 F, 1 F and 10/3 F respectively.
    }
\label{k04}
\end{figure}

\begin{figure}[htbp]
\vspace{-0.5in}
\begin{center}
{\mbox{\epsfig{file=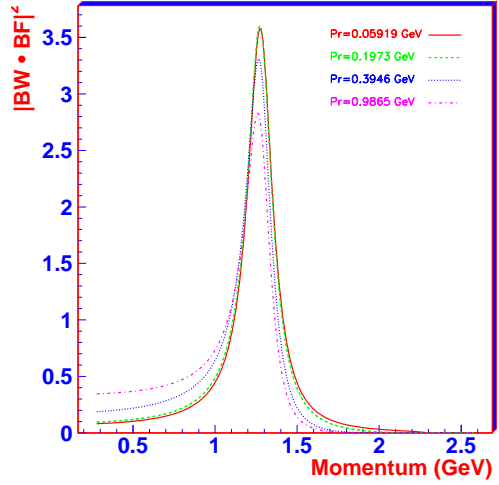,height=2.8in,width=2.8in}}}
\end{center}
\caption[]{The shape of $|BW(s,M,\Gamma) \cdot BF^{(2)}(4,k_2 R_2)|^2$ with
interaction radius $R_0$ 0.2 F, 0.5 F, 1 F and 10/3 F respectively.
    }
\label{k05}
\end{figure}

\section{Interaction Radius}

From figure 4 and figure 5, we could see that the line shape of a resonance
is affected by its centrifugal barrier factor. It is also sensitive to
the value of the resonance's interaction radius. Different values of
the interaction radius will give out different line shapes. Since the line
shape of a resonance is sensitive to its interaction radius, it is possible
for us to measure the interaction radius in high energy experiments.
\\

In  physics analysis, a resonance is traditionally described by the Breit-Wigner
function, which has two independent parameters, the mass and the width of the
resonance. In fact, some information are missing in this treatment.
To describe a resonance, we need at least three parameters, they are the mass,
the width and the interaction radius of the resonance. The mass of the resonance
gives out the central position of the peak, the width gives
out the width of the peak, and the interaction radius gives
out the asymmetry of the resonance's line shape.
\\

In the traditional physics analysis, the interaction radii of all particles
are set to 1 F. (In some physics analysis, the centrifugal barrier factor
is not considered, which corresponds to the case that the interaction radii
for all particles are set to infinity.)
Such treatment will cause large systematic errors to the measurement of
mass, width and branching ratio of the intermediate resonance.
A way to solve this problem is to measure the interaction radius in
physics analysis. It is found that, in partial wave analysis, the
log likelihood function is sensitive to the value of the interaction radius,
and the interaction radius of an intermediate resonance can be well determined
by the technique of the interaction radius scan. In the interaction radius
scan, the mass and width of a resonance is fixed, and the interaction radius
$R$ is the only parameter to be changed. It is found that the log likelihood
function is changed when the parameter $R$ is changed. There is a
maximum of log likelihood function, and the corresponding parameter $R$ gives
out the interaction radius of the resonance. By using this method,
the interaction radii of some resonances are successfully determined
in a analysis on BESII $J/\psi \to \omega \pi \pi$ and
$J/\psi \to K^*(892) K \pi$ data. Old analysis on these data have
already been published\cite{10,11}.
\\

A test is done on a Monte Carlo data. First, a Monte Carlo data is
generated. The data sample consists of 1000
$J/\psi \to \omega X(1270) \to \omega \pi \pi$ events. The interaction
radius of X(1270) is set to 255 MeV. Then we make an interaction radius
scan on this data. The scan curve is shown in Fig. 6. The minimum of the
curve gives out the interaction radius of $X(1270)$ of the Monte Carlo
data. From this test, two basic conclusions can be drawn. One is that the
log likelihood function is sensitive to the value of the interaction
radius. The other is that the magnitude of interaction radius can be
well determined by the interaction radius scan.
\\

\begin{figure}[htbp]
\vspace{-0.5in}
\begin{center}
{\mbox{\epsfig{file=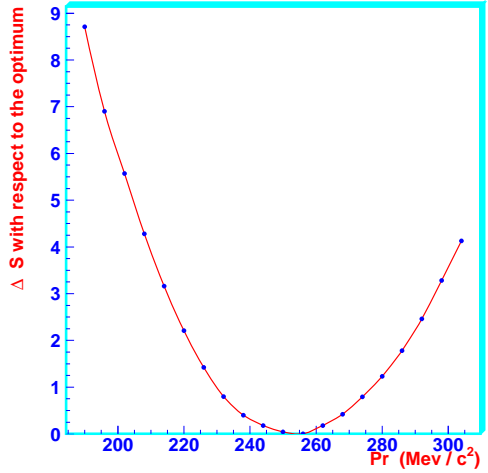,height=2.8in,width=2.8in}}}
\end{center}
\caption[]{Interaction radius scan on a Monte Carlo data. The interaction radius
    of X(1270) is set to 255 MeV.    }
\label{k06}
\end{figure}

\section{Discussions and Results}

In this paper, the theoretically formula used in physics analysis
of $J/\psi \to \omega \pi \pi$ and $J/\psi \to K^*(892) K \pi$
channel is given. In the PWA analysis
of these channels, it is found that our traditional treatment that
all interaction radii are set to 1 F will cause quite large
systematic uncertainties. The systematic uncertainties cause by it
is too large to be neglectable. A way to solve this problem is to
measure the interaction radius in physics analysis.
\\

The interaction radius $R$ of a resonance is the size of the resonance.
It is an important physical quantity to describe a resonance. But for
a long time, we do not know how to measure it in the experiments. Up to now,
we still have no experimental measurements on it. According to our analysis
on BESII data, we found that the interaction radius of a resonance can
be well determined in the PWA analysis. So, we can systematically measure
the interaction radiuses of mesons and baryons  based on BESII and BESIII data.
It open a new research field of high energy experimental physics.
\\

In the traditional treatment, a resonance is describe by two parameters,
the mass and the width. Now, we know that, such treatment is incomplete.
we need a third parameter to describe the resonance. This new parameter
is the interaction radius of the resonance, which gives out the asymmetry
of the resonance's line shape.
\\

\end{document}